# The Prandtl Plus Scaling Failure and its Remedy

By David W. Weyburne[1]
*Air Force Research Laboratory, 2241 Avionics Circle, Wright-Patterson AFB, OH 45433, USA*

## Abstract

The Prandtl Plus scaling parameters are reexamined theoretically. The flow governing equations approach is used to examine similarity issues of the inner region of wall-bounded turbulent flows as well as laminar flow cases. It is found that the Prandtl Plus parameters are in fact similarity scaling parameters for laminar sink flow but NOT the more general laminar Falkner-Skan boundary layer flows. For turbulent boundary layer flows along a wall, it is found that only turbulent sink flows show similarity with the Prandtl Plus scaling parameters. This is diametrically opposed to the accepted notion that the Prandtl Plus parameters work for every wall-bounded turbulent flow. To correct the problem with the Prandtl Plus parameters, we introduce a new set of scaling parameters that work for the Falkner-Skan boundary layer flows and satisfy the relevant part of the flow governing equations approach to similarity.

## 1. INTRODUCTION

One of the most fundamental concepts in fluid mechanics is to analyze experiment observables using dimensional analysis with the intent of finding scaling parameters that render the scaled observable from different stations along the flow to appear to be similar. Similarity of the velocity profile formed by fluid flow along a wall is one of those fundamental observables. For 2-D flows along a wall, velocity profile similarity is defined as the case where two velocity profiles taken at different stations along the wall differ only by simple scaling parameters in $y$ and $u(x,y)$, where $y$ is the normal direction to the wall, $x$ is the flow direction, and $u(x,y)$ is the velocity parallel to the wall in the flow direction. Similarity solutions of the flow governing equations are well known for laminar flow. Turbulent flow similarity is problematic. Since the equations for turbulent flows do not admit to exact similarity solutions, the community has long sought to establish their possible existence by searching for scaling parameters that collapse experimental velocity profile datasets to a single curve.

While the search for scaling in the outer region of the turbulent boundary layer region is an ongoing affair, there has been no disagreement about the proper scaling for the region of the turbulent boundary layer near the wall where viscosity is important. There has been universal consensus that the parameters proposed by Prandtl [1] are correct. The scaled Prandtl "Plus" scaling parameters are the friction velocity $u_\tau$ and the length scaling parameter is $\nu/u_\tau$, where $\nu$ is the kinematic viscosity. However, despite almost a century of effort, there is no theoretical proof that that the Prandtl Plus parameters are similarity parameters for any part of a 2-D wall-bounded turbulent boundary layer flow or even the simpler 2-D wall-bounded laminar boundary layer flow. Herein we start by using the flow governing equations approach to similarity to show that the Prandtl Plus scalings are in fact similarity scaling parameters for laminar sink flow but NOT the more general laminar Falkner-Skan [2] boundary layer flows. We then show that the only wall-bounded turbulent flow that shows similarity with the Prandtl Plus parameters is for the turbulent sink flow case. This is diametrically opposed to the accepted notion that the Prandtl Plus parameters work for every wall-bounded turbulent flow. To correct the problem with the Prandtl Plus parameters, we introduce a new set of similarity scaling parameters that work for the Falkner-Skan boundary layer flows and are also compatible with the relevant part of the flow governing equations approach to similarity.

[1]Email: dweyburne@gmail.com

## 2. Test Cases

The intent of this section is to determine if similarity is present when we apply the Prandtl Plus parameters to a few well-known wall-bounded flow cases where viscosity plays an important role in the structure of the flows boundary layer. We start off examining two laminar flow cases and then the general turbulent flow case.

### 2.1 Laminar Sink Flow

First, we consider laminar flow in a converging 2-D channel, also called sink flow. Take the velocity in the flow direction as $u(x,y)$. The flow is in the minus $x$-direction along the wall towards the sink at $x$=0 and the $y$-direction is perpendicular to the bottom channel. The length and velocity similarity scaling parameters for this flow are well known [3]. The similarity velocity scaling parameter is given by

$$u_s(x) = -\frac{Q}{x} \quad , \tag{1}$$

where $Q$ is the fluid flux and the similarity length scale is given by

$$\delta_s(x) = \sqrt{\frac{-\nu x}{u_s}} = x\sqrt{\frac{\nu}{Q}} \quad . \tag{2}$$

The velocity $u(x,y)$ has a known analytical solution [3] given by

$$u(x,y) = -\frac{Q}{x}\left[3\tanh^2\left(\frac{y}{\sqrt{2}\delta_s(x)} + \tanh^{-1}\sqrt{\frac{2}{3}}\right) - 2\right] \quad . \tag{3}$$

The friction velocity can be derived from the derivative of the velocity evaluated at the wall,

$$\begin{aligned}
\left.\frac{du}{dy}\right|_{y=0} &= \frac{d}{dy}\left[-\frac{3Q}{x}\tanh^2\left(\frac{y}{\sqrt{2}\delta_s(x)} + \tanh^{-1}\sqrt{\frac{2}{3}}\right) - 2\right]_{y=0} \\
\left.\frac{du}{dy}\right|_{y=0} &= -\frac{Q}{x}\left[\frac{1.154}{\delta_s(x)}\right] = -\frac{Q}{x}\frac{1.154}{x\sqrt{\nu/Q}} \\
\left.\frac{du}{dy}\right|_{y=0} &= \frac{1.154}{x^2}\sqrt{\frac{Q^3}{\nu}} \quad ,
\end{aligned} \tag{4}$$

where 1.154 is the rounded value of the exact constant. Using the definition for the friction velocity, this means that for laminar sink flow we have

$$\left.\frac{du}{dy}\right|_{y=0} = \frac{u_\tau^2}{\nu} = \frac{1.154}{x^2}\sqrt{\frac{Q^3}{\nu}} \Rightarrow u_\tau = \frac{1.074}{x}\left(\nu Q^3\right)^{1/4} \quad , \tag{5}$$

By inspection of Eqs. 1, 2, and 5, it is confirmed that the ratios $u_\tau/u_s$ and $(\nu/u_\tau)/\delta_s$ are constants from station to station as required for similarity. Therefore, the Prandtl Plus scaling parameters are similarity scaling parameters for laminar sink flow.

## 2.2 Laminar Flow along a Wall

For a 2-D, incompressible, constant property fluid, the Prandtl laminar boundary layer approximation for the *x*-component momentum balance is given by

$$u\frac{\partial u}{\partial x} + v\frac{\partial u}{\partial y} \cong -\frac{1}{\rho}\frac{\partial p}{\partial x} + \nu\frac{\partial^2 u}{\partial y^2} \quad , \tag{6}$$

where $\rho$ is the density and *p* is the pressure. The velocity *v* is the velocity perpendicular to the wall (*y*-direction). To solve Eq. 6 while ensuring mass conservation, a stream function is introduced which assumes the velocities can be expressed as a separable product of *x* and *y* functionals. Thus, we assume that a stream function $\psi(x,y)$ exists such that

$$\psi(x,y) = \delta_s(x)u_s(x)f(\eta) \quad , \tag{7}$$

where $f(\eta)$ is a dimensionless function that, for similarity, can only depend on the scaled *y*-position, $\eta$, given by

$$\eta = \frac{y}{\delta_s(x)} \quad . \tag{8}$$

The stream function must satisfy the conditions

$$u(x,y) = \frac{\partial\psi(x,y)}{\partial y}, \qquad v(x,y) = -\frac{\partial\psi(x,y)}{\partial x} \quad . \tag{9}$$

For the pressure gradient, we assume the Bernoulli's equation applies given by

$$-\frac{1}{\rho}\frac{\partial p}{\partial x} = u_e\frac{\partial u_e}{\partial x} \quad , \tag{10}$$

where $u_e(x)$ is the velocity at the boundary layer edge.

By non dimensionalizing Eq. 6 using Eqs. 7-10, the *x*-momentum equation reduces to

$$f''' + \alpha ff'' + \beta\left(1 - f'^2\right) = 0 \quad , \tag{11}$$

where the primes indicate differentiation with respect to $\eta$ and where

$$\alpha = \frac{\delta_s^2}{\nu}\frac{du_s}{dx} + \frac{\delta_s u_s}{\nu}\frac{d\delta_s}{dx} \quad \text{and} \quad \beta = \frac{\delta_s^2}{\nu}\frac{du_s}{dx} \quad , \tag{12}$$

(see Appendix A). Similarity of the velocity profile from station to station requires that the Falkner-Skan parameters $\alpha$, $\beta$, and $u_e/u_s$ be constants.

Falkner and Skan [2] determined that similar solutions are obtained for Eqs. 11-12 if one assumes an analytical functional form for $\delta_s(x)$ and $u_s(x)$ of the type

$$\delta_s(x) = \sqrt{\frac{2\nu}{a_s(m+1)}}\left(x - x_0\right)^{(1-m)/2} \quad \text{and} \quad u_s(x) = a_s\left(x - x_0\right)^m \quad , \tag{13}$$

where $a_s$, $x_0$, and $m$ are constants. With these scaling parameters, Eq. 12 reduces to $\alpha = 1$ and $\beta = 2m/(m+1)$. With the similar solution and the definition of the friction velocity in hand, it is straightforward to show that

$$\begin{aligned}
\left.\frac{du}{dy}\right|_{y=0} &= \left[\frac{d\{u_s f'\}}{d\eta}\frac{d\eta}{dy}\right]_{\eta=0} = \frac{u_\tau^2}{\nu} \\
\left.\frac{du}{dy}\right|_{y=0} &= \left[u_s f''(\eta)\frac{1}{\delta_s}\right]_{\eta=0} = \frac{u_\tau^2}{\nu} \\
\left.\frac{du}{dy}\right|_{y=0} &= f''(0)\frac{u_s}{\delta_s} = \frac{u_\tau^2}{\nu} \quad ,
\end{aligned} \tag{14}$$

which means

$$\begin{aligned}
u_\tau(x) &= \sqrt{\nu f''(0)}\sqrt{\frac{u_s}{\delta_s}} = \sqrt{\nu f''(0)}\sqrt{\frac{a_s(x-x_0)^m}{\sqrt{\frac{2\nu}{a_s(m+1)}}(x-x_0)^{(1-m)/2}}} \\
u_\tau(x) &= a_\tau(m+1)^{\frac{1}{4}}(x-x_0)^{\frac{3m-1}{4}} \quad ,
\end{aligned} \tag{15}$$

where $a_\tau$ is a constant. Velocity profile similarity using the Prandlt Plus scaling parameters requires that $u_\tau/u_s$ be a constant. From Eqs. 13 and 15, it is evident that only the $m$=-1 **sink flow case** satisfies the condition for similarity. Hence for laminar Falkner-Skan flows, the Prandtl Plus scaling parameters are **NOT** similarity scaling parameters (except for sink flow).

## 2.3 Turbulent Flow along a Wall

This section is an expanded consideration of our previous work [4]. Consider two-dimensional, incompressible, turbulent flow along a wall. Using the same notations from above, the Prandtl boundary layer approximation for the $x$-component of the momentum balance is usually expressed as

$$\bar{u}\frac{\partial \bar{u}}{\partial x} + \bar{v}\frac{\partial \bar{u}}{\partial y} + \frac{\partial}{\partial x}\left\{\overline{\tilde{u}^2} - \overline{\tilde{v}^2}\right\} + \frac{\partial}{\partial y}\left\{\overline{\tilde{u}\tilde{v}}\right\} \cong -\frac{1}{\rho}\frac{\partial p}{\partial x} + \nu\frac{\partial^2 \bar{u}}{\partial y^2} \quad , \tag{16}$$

where we have used the standard Reynolds decomposition to express the velocities in terms of the average velocities $\bar{u}$ and $\bar{v}$ and the fluctuating components as $\tilde{u}$ and $\tilde{v}$. The momentum balance equation can be put in dimensionless form in the same way that the laminar flow case was done. It turns out that whether one employs a standard stream function $\psi(x,y)$ like in Eq. 7 or a defect profile-based stream function, putting Eq. 16 in dimensionless form results in the same $\alpha$ and $\beta$ terms [5]. These terms are also identical to the laminar flow case versions given by Eq. 12. What changes for the turbulent case is that we have additional terms due to the Reynolds shear stress that also must be constant for similarity. Thus, for the turbulent boundary layer case Eq. 16, the $x$-momentum equation, reduces to

$$f''' + \alpha f f'' + \beta\left(1 - f'^2\right) + \{\text{reduced Reynolds Stress terms}\} = 0 \quad . \tag{17}$$

where

$$\alpha = \frac{\delta_s^2}{\nu}\frac{du_s}{dx} + \frac{\delta_s u_s}{\nu}\frac{d\delta_s}{dx} \quad \text{and} \quad \beta = \frac{\delta_s^2}{\nu}\frac{du_s}{dx} \quad . \tag{18}$$

Similarity of the velocity profile from station to station requires that the Falkner-Skan parameters $\alpha$, $\beta$, and $u_e/u_s$ must be constants. The Reynolds Stress terms add additional constraints but these additional constraints do not change the fact that $\alpha$, $\beta$, and $u_e/u_s$ must be constant for similarity to be present.

The search for turbulent boundary layer scaling parameters $u_s(x)$ and $\delta_s(x)$ has been the subject of much research since exact mathematical solutions are not possible. However, there has not been any debate that the Prandtl Plus scaling parameters are the correct scaling for the inner region of the turbulent boundary layer. If one assumes that the Prandtl Plus scaling parameters are the correct scaling for the inner turbulent boundary layer region for flow along a wall, then the velocity scaling parameter is given by

$$u_s(x) = u_\tau(x) \quad , \tag{19}$$

and the length scale is given by

$$\delta_s(x) = \frac{\nu}{u_\tau(x)}, \tag{20}$$

which means that the parameter $\beta$ is given by

$$\begin{aligned} \beta &= \frac{\delta_s^2}{\nu}\frac{du_s}{dx} \\ \beta &= \frac{\nu}{u_\tau^2}\frac{du_\tau}{dx} \quad . \end{aligned} \tag{21}$$

The $\alpha$ parameter becomes

$$\begin{aligned} \alpha &= \frac{\delta_s^2}{\nu}\frac{du_s}{dx} + \frac{\delta_s u_s}{\nu}\frac{d\delta_s}{dx} \\ \alpha &= \beta + \frac{1}{\nu}\nu\frac{d\{\nu/u_\tau\}}{dx} \\ \alpha &= \frac{\nu}{u_\tau^2}\frac{du_\tau}{dx} - \frac{\nu}{u_\tau^2}\frac{du_\tau}{dx} \\ \alpha &= 0 \quad . \end{aligned} \tag{22}$$

Similarity requires that the parameters $\alpha$, $\beta$, and $u_e/u_s$ be constant from station to station. The $\alpha$ term is a constant. If $\beta$ is a constant, then the differential equation given by Eq. 21 has a solution given by

$$u_\tau(x) = -\frac{\nu}{\beta x} \quad . \tag{23}$$

The fact that $u_e/u_s$ must be constant means that $u_e/u_\tau$ must be a constant (the Rotta [6] constraint), implying that $u_e$ also must behave as $1/x$. This type of flow corresponds to sink flow. Therefore, for turbulent flow along a wall, the ONLY situation for which the Prandtl Plus parameters are similarity scaling parameters is for the turbulent sink flow case.

The momentum equation given by Eq. 16 is applicable for a turbulent flow with a pressure gradient present in the flow direction. For the zero-pressure gradient (ZPG) case, the dimensionless momentum equation reduces to $\beta = 0$ and $\alpha = 1$ in Eq. 17. It is easily verified that the Prandtl Plus parameters do NOT satisfy these $\alpha$ and $\beta$ requirements for ZPG turbulent flow.

### 3. New Inner Region Scaling for 2-D Wall-bounded Turbulent Flow

In the last section, we looked at a few specific flow cases for the Prandtl Plus scaling parameters. It was evident that the Prandtl Plus parameters are not as universally applicable as previously thought. However, the experimental evidence for the Prandtl Plus parameters is extensive. For example, Log Law plots have shown remarkable similarity like-behavior in the near-wall region when plotted with the Prandtl Plus parameters. Therefore, whatever the correct scalings are must be somehow similar to the Prandtl scalings.

One clue as to what might work comes from the above analysis in Eq. 14. Since $f''(0)$ is a constant, let us define a new velocity scaling parameter, which we will call $u_0$, and a new length scaling parameter $\delta_0$, where

$$\frac{u_0}{\delta_0} \propto \left.\frac{du}{dy}\right|_{y=0} = \frac{u_\tau^2}{\nu} = \frac{\tau_w}{\rho\nu} \quad \Rightarrow \quad \delta_0 = c\frac{\nu u_0}{u_\tau^2} \quad , \tag{24}$$

where *c* is a proportionality constant. The scaling ratio is directly proportional to the wall shear stress $\tau_w$. To ensure at least part of the flow governing equations criteria for similarity are met, we require the Falkner-Skan $\alpha$ and $\beta$ terms (Eq. 18) to be constants. Using the new length scale $\delta_0$ and the velocity scale $u_0$ as the similarity parameters we have

$$\beta = \frac{\delta_0^2}{\nu}\frac{du_0}{dx} = \frac{1}{\nu}\left(c\frac{\nu u_0}{u_\tau^2}\right)^2\frac{du_0}{dx} = \nu c^2\frac{u_0^2}{u_\tau^4}\frac{du_0}{dx} \tag{25}$$

$$c^2 u_0^2\frac{du_0}{dx} = \frac{\beta u_\tau^4}{\nu} \quad .$$

It is not possible to separate out $\delta_0$ and $u_0$ as stand-alone parameters. Eqs. 24 and 25 (with $\alpha$ and $\beta$ as constants) are the formal definitions of the new scaling parameters. These new scaling parameters, therefore, satisfy at least part of the flow governing equations approach to similarity, the part which the Prandtl Plus scaling's does not satisfy.

To explore these new parameters further, we need to know the functional form for the friction velocity $u_\tau$ for turbulent flows. The functional form is of course not known in general. In this context, one possibility is to assume Falkner-Skan [2] power-law type boundary layer behavior. Therefore, we assume analytical functions of the type

$$u_0(x) = a_0(x - x_0)^m \quad \text{and} \quad u_\tau(x) = a_\tau(x - x_0)^p \quad , \tag{26}$$

where $a_0$, $a_\tau$, $x_0$, *m*, and *p* are constants. Substituting into Eq. 25 we have

$$c^2 u_0^2 \frac{du_0}{dx} \;=\; \frac{\beta u_\tau^4}{\nu} \tag{27}$$

$$nc^2 a_0^3 \left(x-x_0\right)^{2m} \left(x-x_0\right)^{m-1} \;=\; \frac{\beta}{\nu} a_\tau^4 \left(x-x_0\right)^{4p}$$

$$nc^2 a_0^3 \left(x-x_0\right)^{3m-4p-1} \;=\; \frac{\beta}{\nu} a_\tau^4 \quad ,$$

for flows with non-zero pressure gradients. Similarity requires that $3m$-$4p$-1 = 0 (*i.e.*, $\beta$ must be a constant). Thus, for flows obeying Falkner-Skan power type behavior, the new length and velocity scales are given by

$$u_0\left(x\right) = \; a_0\left(x-x_0\right)^m \quad , \quad \delta_o\left(x\right) = c\frac{\nu u_0}{u_\tau^2} = c\frac{\nu a_0}{a_\tau^2}\left(x-x_0\right)^{(1-m)/2} \quad , \tag{28}$$

$$\text{and} \qquad u_\tau\left(x\right) = a_\tau\left(x-x_0\right)^{\frac{3m-1}{4}} \quad ,$$

which are the same as Falkner-Skan's original Eqs. 13 and 15. For this set of parameters, it can be shown that $\alpha$ and $\beta$ are constants. The flow governing equations similarity method requires that $u_e/u_0$ must also be a constant which means that $u_e(x)$ must also follow an analytical function similar to $u_0$ in Eq. 28.

A special case of the analysis is the zero-pressure gradient (ZPG) case. The momentum equation for this case reduces to $\beta = 0$ and $\alpha = 1$. Since $\beta$ is zero, we cannot use Eq. 25 to define $\delta_0$ and $u_0$ but we must switch over to $\alpha$ in Eq. 18. Doing so, it is straightforward to show that $m$=0. Thus for the ZPG case, we have

$$u_0\left(x\right) = \; \text{constant}, \quad \delta_o\left(x\right) = \left(x-x_0\right)^{1/2} \quad , \quad \text{and} \quad u_\tau\left(x\right) = a_\tau\left(x-x_0\right)^{-1/4} \quad . \tag{29}$$

An important aspect of these new parameters using Eqs. 24-29 is that they DO satisfy the Falkner-Skan flow similarity constraints as opposed to the Prandtl Plus parameters, which do NOT.

The new parameters and the Prandtl Plus parameters will behave similarly when the friction velocity has the same $x$-behavior as the new velocity parameter (Eq. 28). This will happen when

$$\left(x-x_0\right)^m \;\approx\; \left(x-x_0\right)^{(3m-1)/4} \;\Rightarrow\; m \;=\; -1 \quad . \tag{30}$$

Hence both the Prandtl Plus parameters AND the new parameters $\delta_0$ and $u_0$ are similarity scaling parameters for laminar and turbulent sink flows. Thus the new parameters work for the Falkner-Skan power-law fluids and sink flow fluids.

## 4. Experimental

The experimental verification of the new inner region length scale is not straightforward. The problem is that experimentally, the friction velocity is almost universally determined [7] using the Clauser [8] chart method. The Clauser chart method infers that there is a near-wall region in which the velocity profile behaves logarithmically when plotted using the Prandtl Plus scaling parameters. Experimentally, it is known that this method compares reasonably well to other experimental measures of the friction velocity. However, the theoretical arguments above indicate that the Prandtl Plus scaling parameters are not the correct parameters for general wall-bounded turbulent flows. Hence it is possible that even if the extracted friction velocity data compares well with other techniques, it will be biased in the sense that plot comparisons will tend to favor the Prandtl Plus parameter plots since the friction velocity data was extracted assuming Prandtl Plus parameters were valid. Therefore, one needs to use datasets in which the friction velocity was determined independently.

It is a challenge to find and obtain datasets in which the friction velocity was determined independently. The dataset from Skåre and Krogstad [9] comes close to this requirement. In their dataset, several different methods were used to obtain the friction velocity [9]. In Fig. 1 the friction data is plotted against the distance along the wall. A good fit to Eq. 28 (red line) using the experimental data (■) yielded $m$=-0.133. A $x_0$ = 1.74 value was obtained from the displacement thickness linear fit. In working with experimental equilibrium-type turbulent boundary layers, we almost always see that both the displacement thickness and the momentum thickness are linear functions of $x$ with the same $x_0$ intercept, in line with Townsend's conjecture [10]. By using this $x_0$ value we are insuring thickness and velocity scaling parameters are compatible with the displacement thickness and momentum thickness. This $x_0$ and the fitted $m$-value were then used in Eq. 28 to generate the $\delta_0(x)$ and $u_0(x)$ data. The constants $c$, $a_0$, $a_\tau$, and $\nu$ in Eq. 28 are simply set to one.

In Fig. 2 we present side-by-side plots of the Skåre and Krogstad [9] profiles using the Prandtl Plus scaling's and the new inner region scaling's. The plots reveal that the differences are not large between the two scalings in the near-wall region. While by no means definitive, this positive result indicates that there is a justifiable reason to explore this new scaling approach further.

The next step was to determine whether the new parameters would produce similar-type behavior in the inner region of a "typical" dataset. For this purpose, we choose one of Clauser's original datasets [8]. The friction velocity data is presented in Fig. 3. The red line in Fig. 3 is the red square Clauser data fitted to Eq. 28 with $m$=-0.333, $x_0$=-1.26. Using the red square reported values, the Prandtl Plus parameters produce the scaled velocity profile plot shown in Fig. 4a. The good overlap in the Log Law region is evident. Next, we generated the $\delta_0(x)$ and $u_0(x)$ parameters by hand fitting the $m$-value in Eq. 28 using a spreadsheet (the $x_0$ obtained as above). The $m$-value was adjusted to give the best overlap in the inner region of the scaled profile plot. The resulting friction velocity is shown as the blue line in Fig. 3 and the resulting scaled profiles are displayed in Fig. 4b. The blue line friction velocity obtained using Eq. 28 used the $a_\tau$ value taken as the average Clauser friction velocity value (the present method has no way of calculating $a_\tau$ independently). Since our intent is to produce similarity in the inner region, the $m$-values were then individually adjusted to show good overlap in the inner region. In Fig. 4c the resulting profiles are shown. The $m$-values were adjusted by only 1-2% to produce the noticeably better fit. The adjusted friction velocity data is shown in Fig. 3 as the blue diamonds using Eq. 28 with the $a_\tau$ value taken as the average Clauser friction velocity value.

We must emphasize that at present there is no way to generate friction velocity data independently using $\delta_0(x)$ and $u_0(x)$ polynomial parameters. There is not an equivalent Log Law at this point for the new parameters. The plots in Fig. 4 only require relative values so in those cases we could have simply set the polynomial constants, $a_0$ and $a_\tau$ in Eq. 28, to any value. The results in Fig. 3 using the $a_\tau$ value taken as the average Clauser friction velocity value was done merely to demonstrate that it is possible to produce similar-like profile behavior in the inner region using the new method with friction velocity data that has **comparable *x*-coordinate** behavior to that produced by the Clauser [8] chart method with the Prandtl Plus parameters.

The next dataset we explored was the ZPG DNS data by Sillero, Jimenez, and Moser [12]. This turbulent flow dataset has the advantage that the friction velocity is known exactly. In Fig. 5 we plot the experimental friction data as the **black line** and the fitted line to Eq. 28 as the red line. For this fit, we again used the $x_0$ value obtained from the displacement thickness linear fit. Although not shown, we also tried fitting the friction velocity to $m$=-1, $m$=0, and $m$=0.753 but the fits are much worse than the red line (why we tried these other $m$-values will become apparent below).

The Prandtl Plus parameter plot in Fig. 6a is a very good fit as is the $m$=0.753 fit shown in Fig. 6b. The results for the $m$=0.753 figure are obtained by hand fitting the $m$-value in Eq. 28 using a spreadsheet (the $x_0$ obtained as above). Note that there is no individual adjustment of the $m$-values as we did for the Clauser case. The scaled profiles displayed in Fig. 6c are obtained using a $m$=0.175 value from the fit to the friction data (Fig. 5). Since this is a ZPG dataset, we should have $m$=0, $u_0$ constant. The scaled profile data using this scaling value is displayed in Fig. 6d. The Prandtl Plus parameters and $\delta_0$ and $u_0$, $m$=0.753 plots are superior to these other $m$-value dataset plots. This apparently anomalous behavior is discussed in the next section.

## 4. Discussion

What is remarkable about the supposed failure of the Prandtl Plus parameters for the general Falkner-Skan flow situation is that it does not seem to be consistent with experimental results for the last sixty years. Surely some groups would have noticed discrepancies in the calculated friction velocity by the Clauser chart method and direct experimental measurements. Instead, what is observed is that in most cases the direct experimental measurements of the skin friction agree reasonably well with the Clauser Chart method with a few exceptions (see discussion in George [7]). We note there may be an explanation for why there have not been major discrepancies leading to a major controversy. It turns out that while **most** wind tunnel experiments may not be normally associated with sink flow, they mimic it in actual experimental detail. Sink flow is flow in a converging/diverging channel. Most wind tunnel experiments actually fall into this category. Consider the zero-pressure gradient (ZPG) flow situation for example. ZPG flows are notoriously difficult to generate in a wind tunnel but the easiest way to induce a ZPG flow is to slightly tilt the floor or ceiling of the wind tunnel by a few degrees in order to cancel out the natural pressure gradient induced by the growing boundary layer. Many FPG and APG experimental flows also fall into this category since the pressure gradients are often created by adjusting the floor/ceiling height along the flow direction. The equilibrium (similarity) flow attempts of Caluser [8], and Skåre and Krogstad [9] fall into this category. It is challenging to find experimental datasets that cannot be considered a type of converging/diverging channel flow. However, it must be recognized that flow over a wing of an airplane is NOT like sink flow but instead is more closely associated with Falkner-Skan type of flow. Therefore, the revelations herein may have significant implications for flow modeling for aircraft design. Further research will reveal if the impact is mostly in our theoretical interpretation or if there are both theoretical and experimental consequences.

Since the Prandtl Plus scaling failure could have serious implications for wall-bounded turbulent flows, it was critical that we verify that this failure was real. For this reason, we decided to include laminar flow results. We reason that if the Prandtl Plus parameters are supposed to describe the boundary layer region where viscosity effects are dominant, then it follows that they should work equally well for wall-bounded laminar flow boundary layers. The laminar flow results are based on the well-known exact solutions to the flow governing equations. The results above are unequivocal; the Prandtl Plus parameters do NOT work for the Falkner-Skan laminar flows. The most important point to be made here is that the laminar flow result is based on the same flow governing equations approach to similarity employed for the turbulent case [5, 6, 10, 11]. This strongly supports the approach that we used for the turbulent flow case. That leaves us in the position that either Prandtl Plus scaling is not applicable for general wall-bounded flows, as we claim above, or that the flow governing equations approach to turbulent similarity [5, 6, 10, 11] are problematic.

By essentially reverse-engineering the Prandtl Plus parameter failings, we were able to introduce a set of similarity scales applicable to the inner region of the wall-bounded turbulent boundary layer velocity profile. For the case where the scaling velocity, $u_s(x)$, and the friction velocity, $u_\tau(x)$, are assumed to be simple power functions of *x*, we showed that these new parameters (Eqs. 24-25) turn out to be identical to the original Falkner-Skan's scaling parameters (Eqs. 13 and 15). The reason, of course, is that we are forcing the Falkner-Skan $\alpha$ and $\beta$ parameters (Eq. 18) to be constants. In contrast, we showed in Section 2.3 above that the only wall-bounded turbulent flow that the Prandtl Plus parameters make $\beta$ a constant is for the sink flow case. If one examines the way the flow governing equations approach to similarity defines $\beta$ (see [5] for example), one sees that the $\beta$ parameter is associated with the viscosity term of the momentum equation. This is precisely the term that dominates the near-wall behavior of fluid flow. Thus, the new method ensures the correct viscosity scaling whereas in general the Prandtl Plus scaling does not.

The new parameters (Eqs. 24-25) were first tested by comparing their fit to experimental datasets which, as we just pointed out, maybe cases of sink-like flow. The plots of the velocity profiles using the two-parameter sets look very similar as is evident in Figs. 2, 4, and 6. The data from Clauser is interesting. In Fig. 4a the Plus parameters result in good overlap in the Log Law region. The results with tweaking the new parameters shown in Fig. 4c give very similar results in this near-wall region. The justification for individually adjusting the *m*-values for this case is that the Clauser friction velocity red square data in Fig. 3 is essentially doing the same thing, i.e. adjusting each friction velocity value to give the best inner region plots (Fig. 4a). To confirm this, we plotted the profiles using the Clauser fitted friction velocity data (Fig. 3, red line) in Fig. 4d. The overlap in the inner region is noticeably poorer than Fig. 4a. Further confirmation comes from examining scattering behavior in Fig. 3. Notice that the blue diamonds have the same scatter behavior to the blue line as the red square data do to the red line. One can only speculate as to whether the scatter is due to forcing the appearance of inner region similarity or whether there are experimental justifications for the scatter.

Overall, the results herein indicate that the new parameter set can produce behavior that is comparable to that obtained by the Prandtl-Plus-based Clauser chart method. At a minimum, this demonstrates that there are other scaling parameters that are at least as good as the Prandtl Plus scaling parameters in terms of inducing similar-like behavior in the inner region of plots of experimental sets of velocity profiles.

The computer-generated ZPG data by Sillero, Jimenez, and Moser [12] plotted in Figs. 5 and 6 is interesting and, if anything, somewhat anomalous. In this case, we know exactly what the friction velocity is for each profile in the dataset. In Fig. 5, the exact friction velocity data is shown as the **black line** and the fitted friction velocity to Eq. 26 ($m$=0.175, $x_0$=-604) is shown as the red line. However, this is a ZPG dataset, so we should have $m$=0. We tried to fit the data with $m$=0, the resulting line (not shown) is a very poor fit to the experimental data. Next, we tried to fit the friction data to the $m$=0.753 result that was obtained by fitting the velocity profile data (Fig. 6b). This friction velocity fit was much worse; the calculated friction velocity increases with $x$ instead of decreasing. Yet the resulting profile plot (Fig. 6b) is a very good fit as is the Prandtl Plus plot (Fig. 6a). It is not clear why this dataset does not behave as a ZPG dataset in this regard. It is also not clear why the Prandtl Plus parameters work so well since the only sink flow-like characteristic is the flow geometry. Finally, it is also not clear why the $m$=0.753 parameter set for $\delta_0$ and $u_0$ works so well compared to $m$=0 or $m$=0.175. At this point, only more research will reveal what exactly is going on in this dataset and verify if the new parameters work as well experimentally as the Prandtl Plus parameters have in the past.

## 5. Conclusion

Using the flow governing equations approach to similarity, it is found that the ONLY situation for which the Prandtl Plus parameters are similarity scaling parameters is for the laminar and turbulent sink flow cases. The theoretical failure of the Prandtl Plus parameters for general turbulent boundary layer flows will necessitate a reexamination of existing theory and practice for wall-bounded turbulent flows. It may be that the biggest consequence is that the theoretical foundation is not as solid as previously thought.

A new set of Falkner-Skan similarity scaling parameters for the inner region of the turbulent boundary layer velocity profile are introduced based on forcing the Falkner-Skan $\alpha$ and $\beta$ terms to be constant. The new parameters work for both laminar and turbulent Falkner-Skan flows. Based on the limited experimental results herein, there is enough to justify further research.

## Acknowledgment

The work was performed on behalf of the Air Force Research Laboratory and the support of AFOSR (Gernot Pomrenke). In addition, the author thanks the experimentalists for making their datasets available for inclusion in this manuscript.

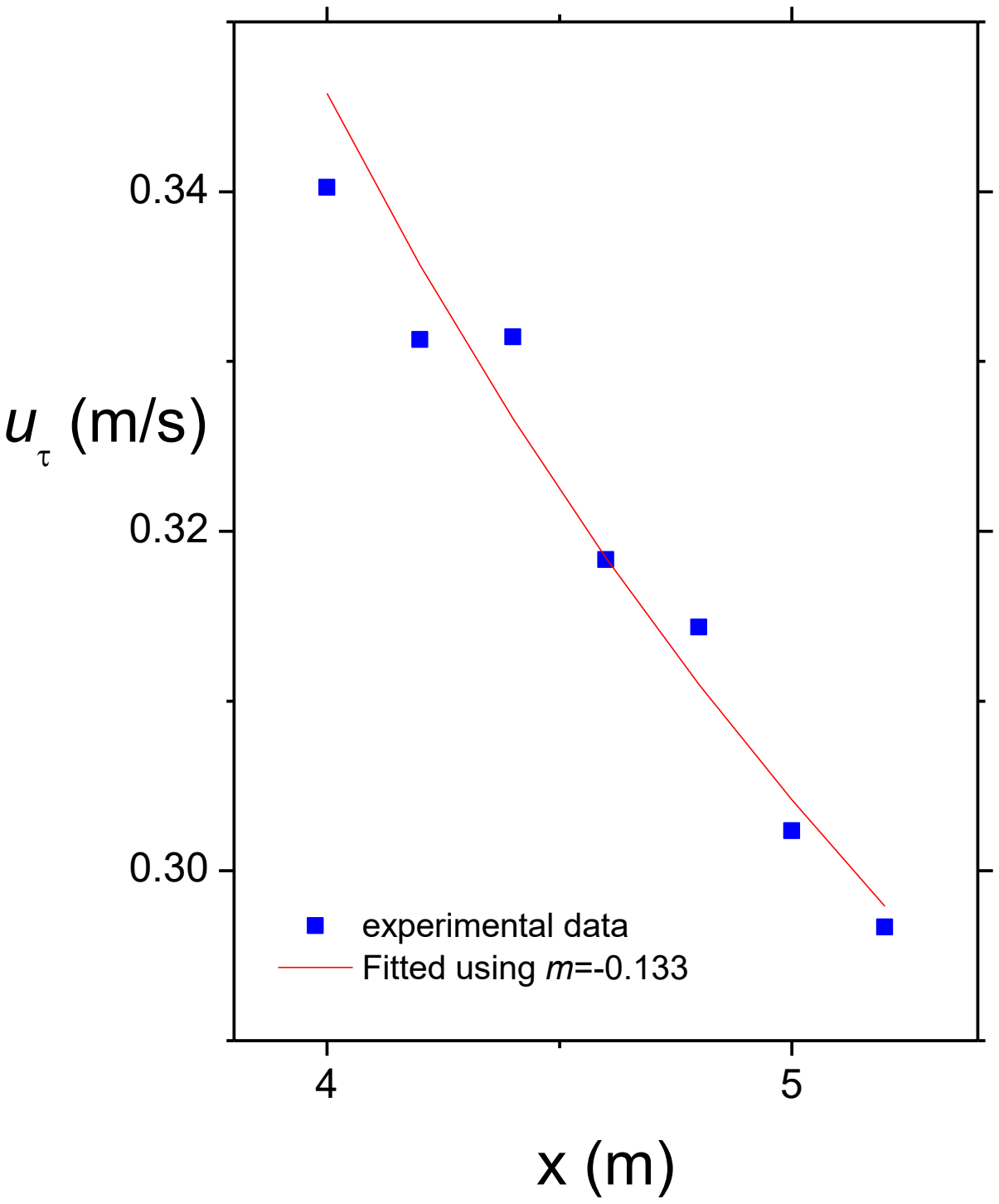

Figure 1: Skåre and Krogstad's [9] friction velocity data (■) and the fitted line (red line) to Eq. 28 plotted against the wall location.

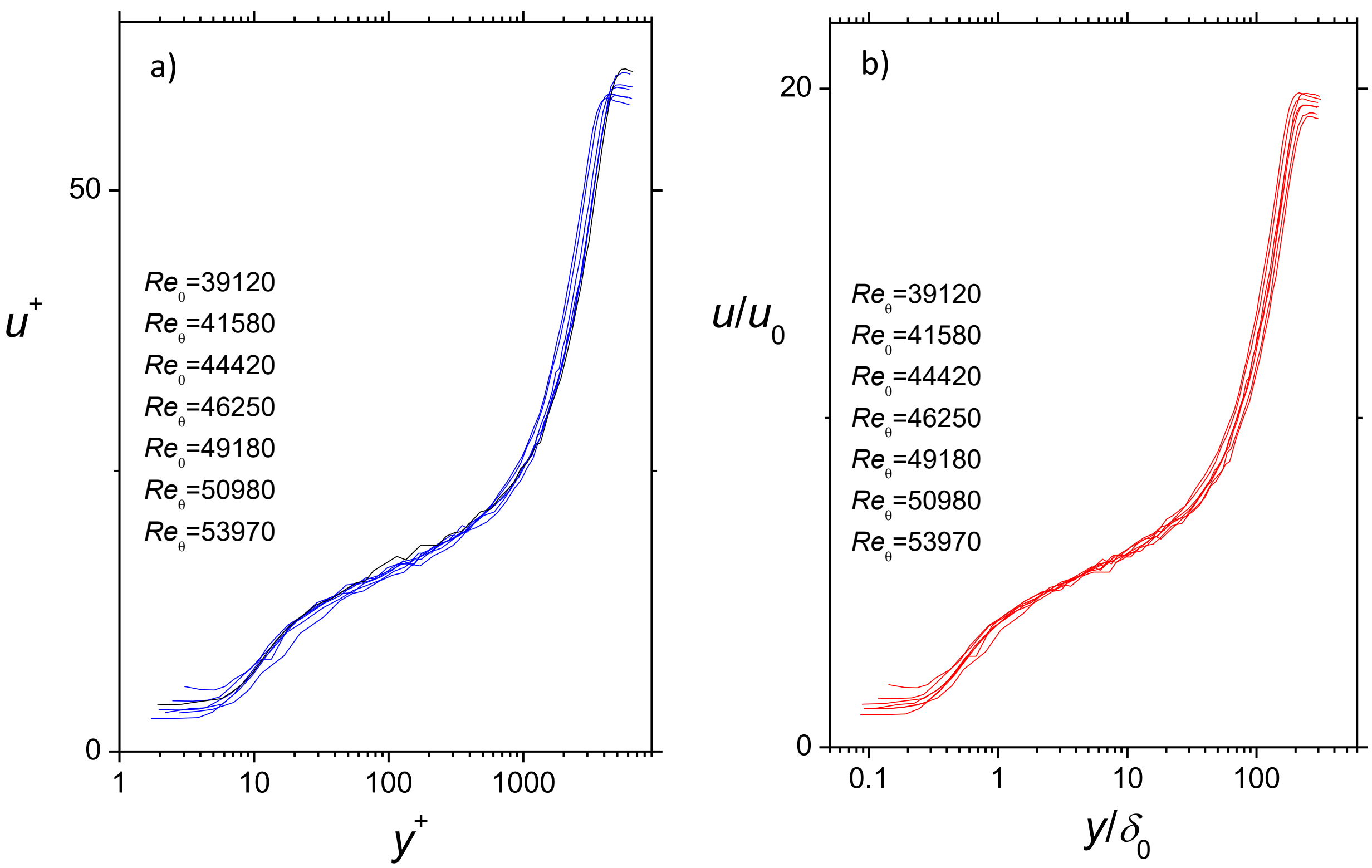

Figure 2: Skåre and Krogstad's [9] profile data plotted a) using $y^+$ and $u^+$, b) using $\delta_0$ and $u_0$. The $Re_\theta$ list identifies which profiles are plotted.

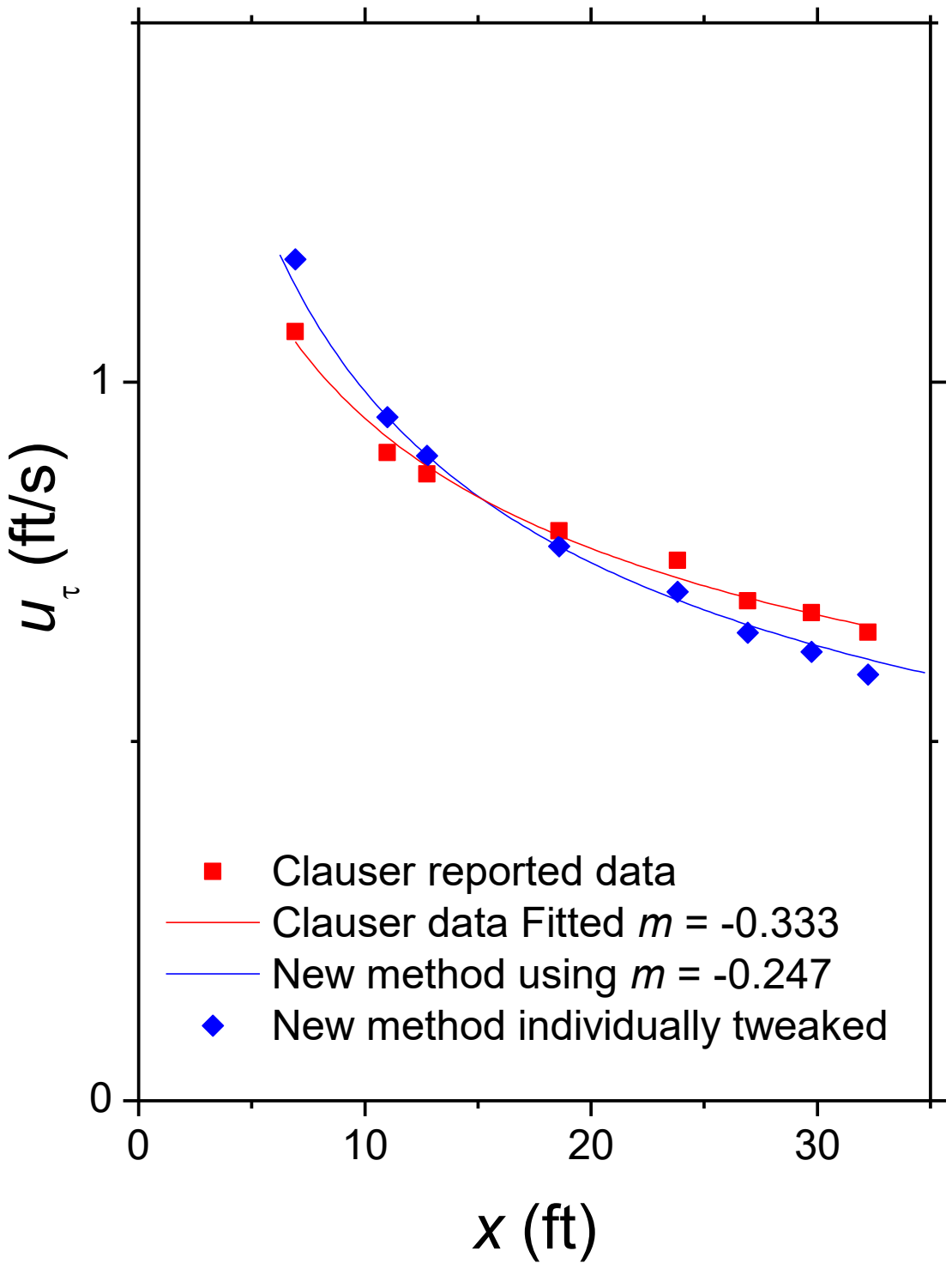

Figure 3: Clauser's [8] friction velocity, $u_\tau$, data (■). The red line is Eq. 28 with $m$=-0.333, $x_0$=-1.26 and the blue line is Eq. 28 with $m$=-0.247, $x_0$=-1.26. The ♦ points are calculated starting with $m$=-0.247 but the $m$-values values are tweaked so that the velocity profile plot show similar-like behavior in the near wall region (see Fig. 4).

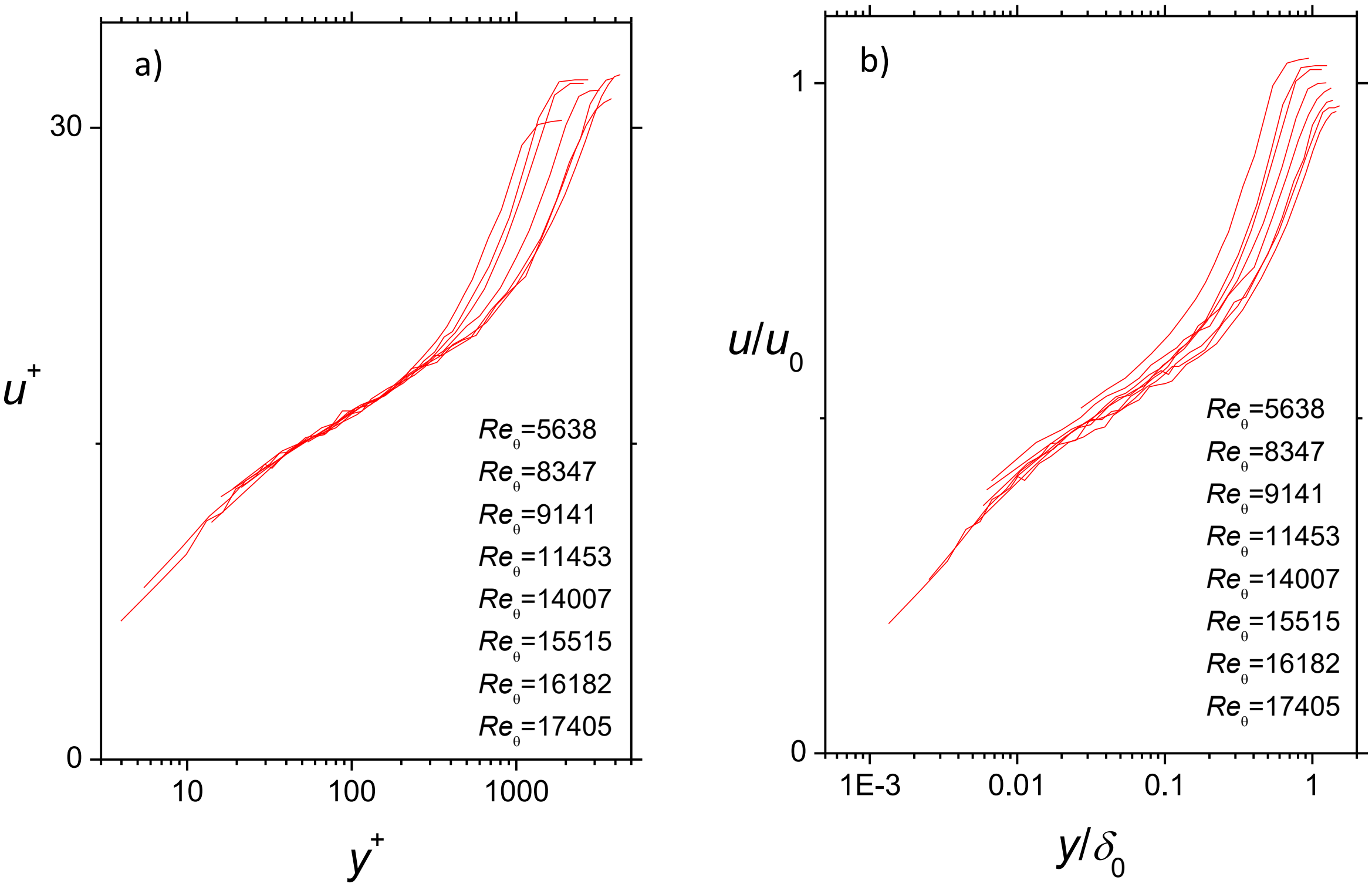

Figure 4: Clauser's [8] scaled profile data plotted a) using $y^+$ and $u^+$, b) using $\delta_0$ and $u_0$ with $m$=-0.247, c) using $\delta_0$ and $u_0$, starting with $m$=-0.247, but individually tweaking the $m$-values to give similar-like behavior near the wall. The resulting tweaked $u_\tau$ values are shown in Fig. 3. In d) the $y^+$ and $u^+$ data is calculated using the Red Line fitted friction velocity data from Fig. 3.

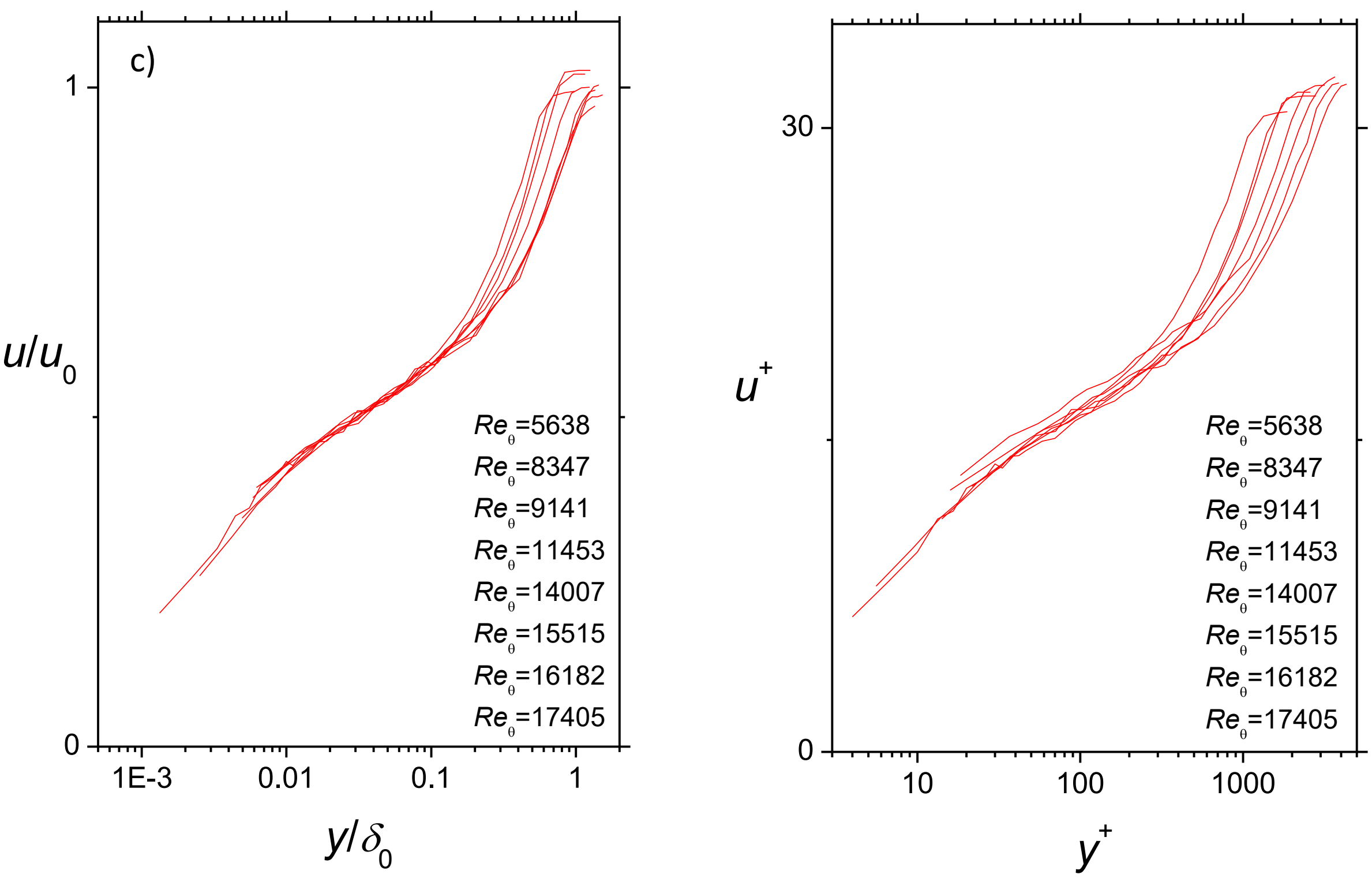

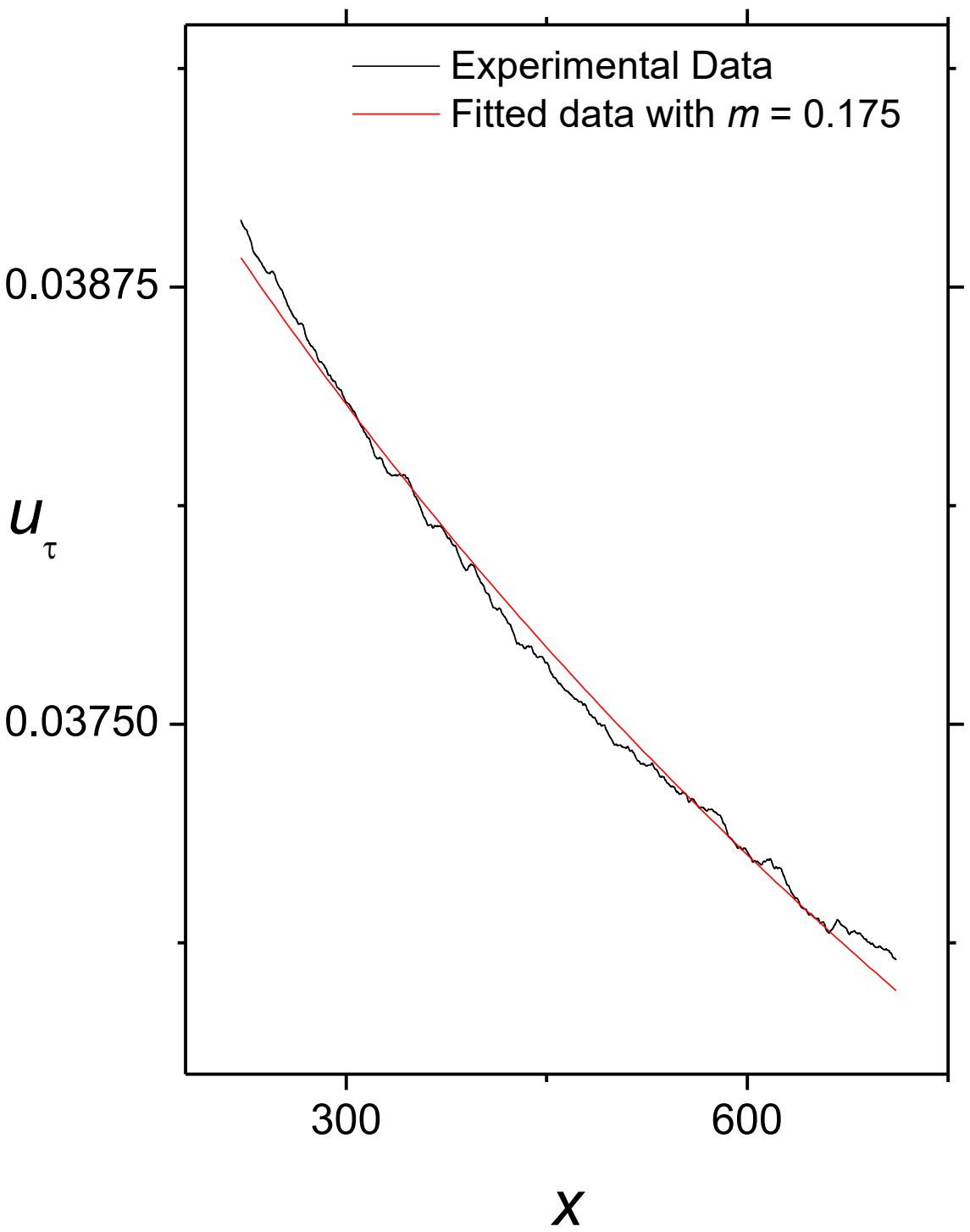

Figure 5: Sillero, Jimenez, and Moser's [12] friction velocity, $u_\tau$, is the **black line** and the red line is the fitted line ($m$ = 0.175, $x_0$ = -604) to Eq. 28.

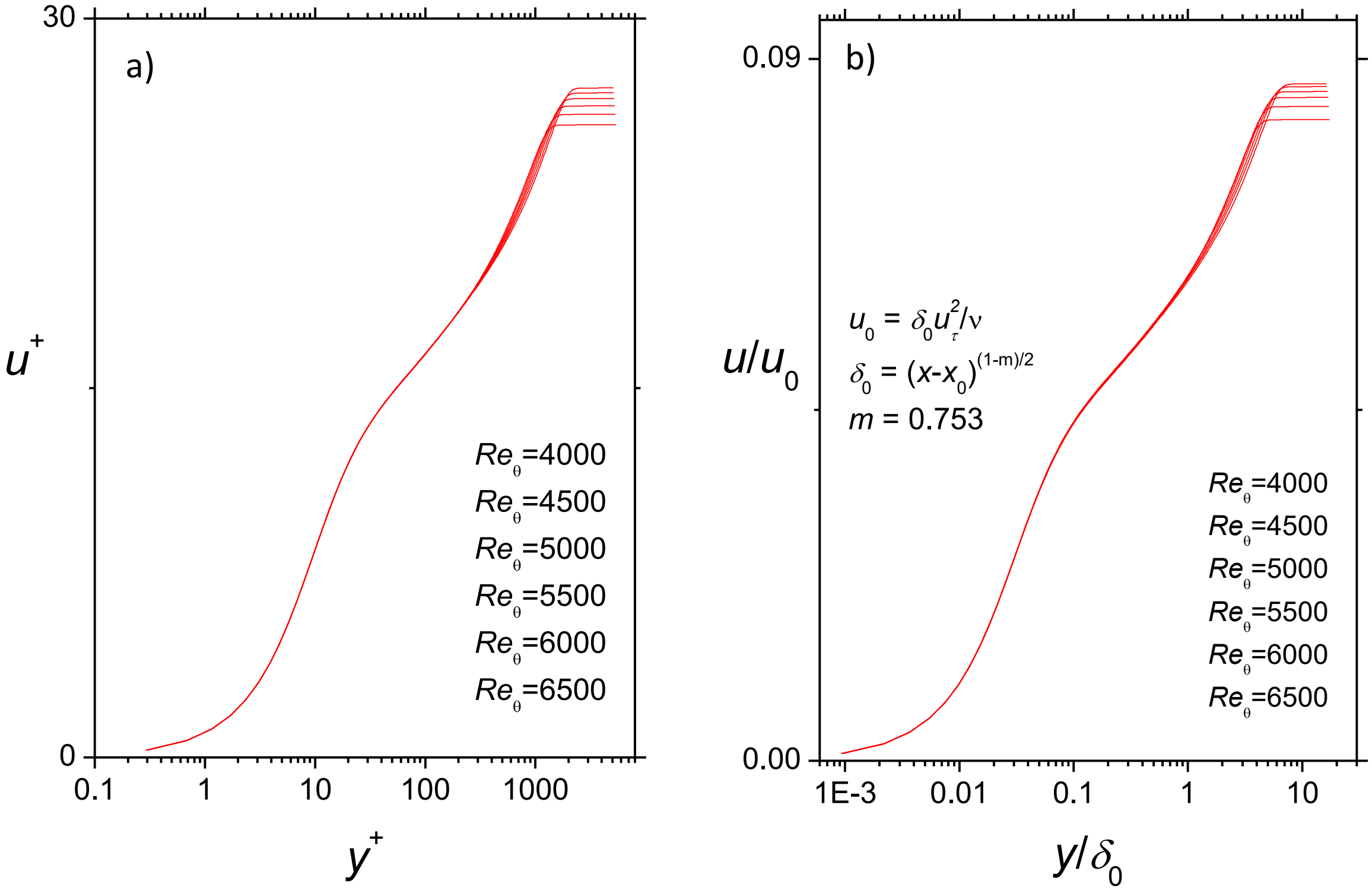

Figure 6: Sillero, Jimenez, and Moser's [12] scaled profile data plotted as: a) using $y^+$ and $u^+$, b) using $\delta_0$ and $u_0$ with $m$=0.753 and $x_0$ = -604, c) using $\delta_0$ and $u_0$ with $m$=0.175 and $x_0$ = -604, and d) using $\delta_0$ and $u_0$ with $m$=0 and $x_0$ = -604.

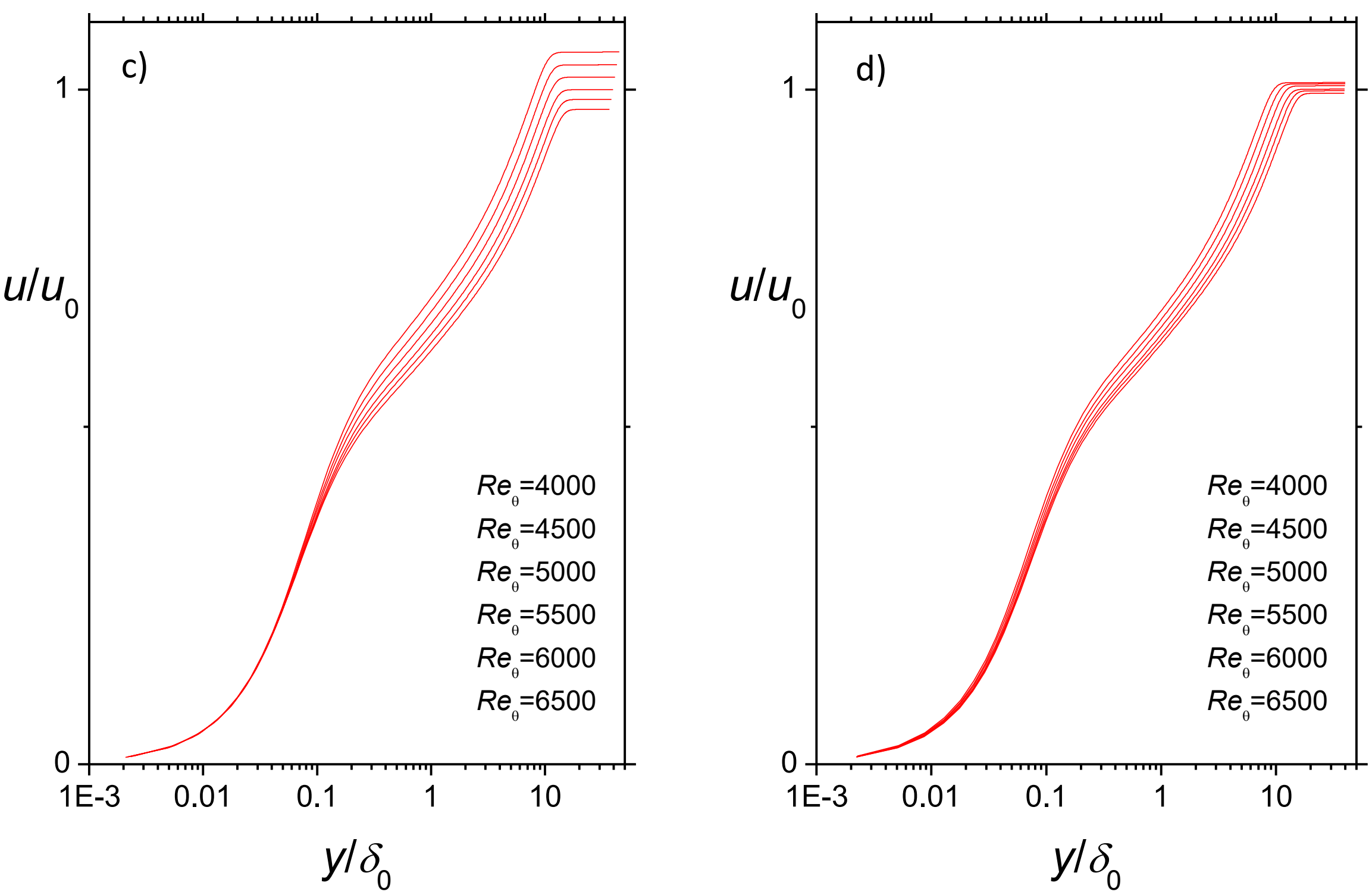

## Appendix A: The Falkner-Skan Equation

The derivation of the dimensionless $x$-momentum equation used above for the Falkner-Skan equations using the stream function given by Eq. 7 is not the standard expression often seen in the literature. Part of the difference is that we denote the velocity profile scaling parameters as $\delta_s(x)$ and $u_s(x)$ which is also somewhat different from what is usually seen in the literature. The stream function version we used herein most closely compares to the version used by Rodgers in his Laminar Flow Analysis book [14] (see his derivations in his Section 3-4). However, given the importance of this derivation to the above analysis, we offer the following comparison to the much more well-known version presented by Schlichting in his Boundary Layer Theory book [3].

To this end, we first summarize the Schlichting expressions. The dimensionless independent variables used by Schlichting are

$$\xi \;=\; \frac{x}{L}, \quad \eta = \frac{\mathrm{Re}^{1/2}}{g(x)}\frac{y}{L} \quad , \tag{A.1}$$

where $L$ is a constant with units of distance and the function $g(x)$ is the unknown thickness-like scaling parameter, and

$$Re \;=\; \frac{LU_\infty}{\nu} \quad , \tag{A.2}$$

where $U_\infty$ is the free stream velocity. Schlichting assumes that a stream function $\psi(x,y)$ exists such that

$$f(\xi,\eta) \;=\; \frac{\psi(x,y)Re^{1/2}}{Lg(x)U(x)} \quad , \tag{A.3}$$

where $U(x)$ is the velocity scaling parameter, $f(\xi,\eta)$ is a dimensionless function, and the stream function $\psi(x,y)$ satisfies the conditions

$$u(x,y) \;=\; \frac{\partial\psi(x,y)}{\partial y}, \qquad v(x,y) \;=\; -\frac{\partial\psi(x,y)}{\partial x} \quad . \tag{A.4}$$

Using this stream function together with the scaling parameters given by Eq. A.1, the reduced Prandtl boundary layer version of the $x$-momentum balance equation (Schlichting's Eq. 8.15) becomes

$$f''' + \alpha f f'' + \beta\left(1 - f'^2\right) \;=\; 0 \quad , \tag{A.5}$$

which is the familiar Falkner-Skan similarity equation for laminar flow on a flat plate with a pressure gradient.

The parameter $\alpha$ as given by Schlichting's Eq. 8.14 is

$$\alpha \;=\; \frac{Lg^2 \dfrac{dU}{dx}}{U_\infty} + \frac{LgU\dfrac{dg}{dx}}{U_\infty} \quad , \tag{A.6}$$

and his $\beta$ is

$$\beta \;=\; \frac{Lg^2}{U_\infty}\frac{dU}{dx} \quad . \tag{A.7}$$

The Falkner-Skan version used in Section 2.2 above uses a different way of expressing the velocity profile scaling parameters and the stream function. In the version used in Section 2.2 above, the length scaling parameter is $\delta_s(x)$ and the velocity scaling parameter is $u_s(x)$. Comparing to Eq. A.1, the equivalent scaling parameters given by Schlichting are

$$\eta = \frac{Re^{1/2}}{g(x)}\frac{y}{L} = \frac{y}{\delta_s(x)} \quad \rightarrow \quad \delta_s(x) \;=\; L\frac{g(x)}{Re^{1/2}} \quad \text{and} \tag{A.8}$$

$$U(x) \;=\; u_s(x) \quad .$$

Now we are in a position to compare the $\alpha$ and $\beta$ terms. The $\alpha$ term given above by Eq. 12 and 18 is

$$\alpha \;=\; \frac{\delta_s^2 \dfrac{du_s}{dx}}{\nu} + \frac{\delta_s u_s \dfrac{d\delta_s}{dx}}{\nu} \quad . \tag{A.9}$$

Using Eq. A.8, this equation reduces to

$$\alpha \;=\; \frac{\left(L\dfrac{g(x)}{Re^{1/2}}\right)^2 \dfrac{dU(x)}{dx}}{\nu} + \frac{\left(L\dfrac{g(x)}{Re^{1/2}}\right)U(x)\dfrac{d\left(L\dfrac{g(x)}{Re^{1/2}}\right)}{dx}}{\nu} \tag{A.10}$$

$$\alpha \;=\; \frac{L^2 g^2}{\nu Re}\frac{dU}{dx} + U\frac{L^2 g}{\nu Re}\frac{dg}{dx}$$

$$\alpha \;=\; \frac{\nu L^2 g^2}{\nu L U_\infty}\frac{dU}{dx} + U\frac{\nu L^2 g}{\nu L U_\infty}\frac{dg}{dx}$$

$$\alpha \;=\; \frac{L g^2}{U_\infty}\frac{dU}{dx} + \frac{L g U}{U_\infty}\frac{dg}{dx} \quad .$$

Compare Eq. A.6 to Eq. A.10; they are the same.

Next consider Schlichting's $\beta$ term given by Eq. A.7. Compare this to the version given in Eqs. 12 and 18 given by

$$\beta \;=\; \frac{\delta_s^2 \dfrac{du_s}{dx}}{\nu} \quad . \tag{A.11}$$

Using Eq. A.8, this reduces to

$$\beta \;=\; \frac{\left(L\dfrac{g(x)}{Re^{1/2}}\right)^2 \dfrac{dU(x)}{dx}}{\nu} \;=\; \frac{\dfrac{L^2 g^2}{Re}\dfrac{dU}{dx}}{\nu} \tag{A.12}$$

$$\beta \;=\; \frac{\nu L^2 g^2}{\nu L U_\infty}\frac{dU}{dx}$$

$$\beta \;=\; \frac{L g^2}{U_\infty}\frac{dU}{dx}$$

Compare Eq. A.7 to Eq. A.12; they are the same. Thus, it is confirmed that the $\alpha$ and $\beta$ expression used by Schlichting are equivalent to the expressions used above.

Now consider the stream function given by Eq. A.3. Using Eq. A8, Eq. A.3 reduces to

$$f(\xi,\eta) \;=\; \frac{\psi(x,y)Re^{1/2}}{Lg(x)U(x)} \;=\; \frac{\psi(x,y)}{\delta_s(x)u_s(x)} \quad . \tag{A.13}$$

Compare Eq. A.13 to Eq. 7 above; they are the equivalent. Hence, the Falkner-Skan stream function and scaling parameters used herein are equivalent to the more widely known version presented by Schlichting [3]. Although the details were not included here, it is easily verified by visual inspection that they are also equivalent to those presented by Rogers [14].